\documentclass[useAMS,usenatbib]{mn2e}

\usepackage{amsmath}
\usepackage{mathtools}
\usepackage{amssymb,amsfonts,amssymb} 
\usepackage{latexsym}
\usepackage{enumerate} 
\usepackage{color,graphicx}
\usepackage{lscape}

\usepackage[letterpaper]{geometry} 
\newcommand{\hpb}{HB13}
\newcommand{\sdss}{\textit{SDSS}}
\newcommand{\galex}{\textit{GALEX}}
\newcommand{\galexmis}{\textit{GALEX-MIS}}
\newcommand{\uvr}{UV$r$2CD}

\title[UVX and binary stars]{Binary stars and the UVX in early type galaxies}
\author[F. Hern\'andez-P\'erez and G. Bruzual]{Fabiola Hern\'andez-P\'erez$^{1}$\thanks{E-mail:
fhernandez@cida.gob.ve} and Gustavo Bruzual$^{2}$\\
$^{1}$Centro de Investigaciones de Astronom\'ia, CIDA, Av. Alberto Carnevalli,  M\'erida, Venezuela. A.P. 264, C.P. 5101\\
$^{2}$Centro de Radioastronom\'ia y Astrof\'isica, CRyA, UNAM, Campus Morelia, Michoac\'an. M\'exico. A.P. 3-72, C.P.58089}
\begin{document}

\date{Accepted 0000 December 15. Received 0000 December 14; in original form 0000 October 11}

\pagerange{\pageref{firstpage}--\pageref{lastpage}} \pubyear{2014}

\maketitle
\label{firstpage}

\begin{abstract} 
We use the \citeauthor*{\hpb} HB13 stellar population synthesis models to study the r\^ole of interacting binary pairs as progenitors of EHB stars.
We assemble a sample of 3417 Early Type Galaxies observed both in the optical (\textit{SDSS}-DR8) and the UV (\textit{GALEX}-GR6).
The galaxies in our sample can be classified according to their position in the colour-colour diagram as UV weak or red sequence galaxies ($\sim 48\%$),
UV strong or UVX galaxies ($\sim 9\%$), and recent star forming galaxies ($\sim 43\%$).
Analysing this sample using the HB13 models for various choices of basic model parameters we conclude that:
\textit{(a)} The UV$r$ colours of UV weak and UV strong galaxies are reproduced by the models as long as the fraction of binary stars is at least 15\%.
\textit{(b)} Higher metallicity models ($Z = 0.02$ and $Z = 0.03$) reproduce the colours of UV weak and UV strong galaxies better than lower $Z$ models.
The $Z = 0.03$ model is slightly bluer than the $Z = 0.02$ model in the UV strong region, indicating a weak relationship between UVX and $Z$.
\textit{(c)} The strength of UVX increases with age in the model population.
This is at variance with the results of other models that include binary stars as progenitors of EHB stars.
\end{abstract}


\section{Introduction}
\label{sec:intro}

\begin{figure}
\centering
\includegraphics[width=1.04\columnwidth,angle=0]{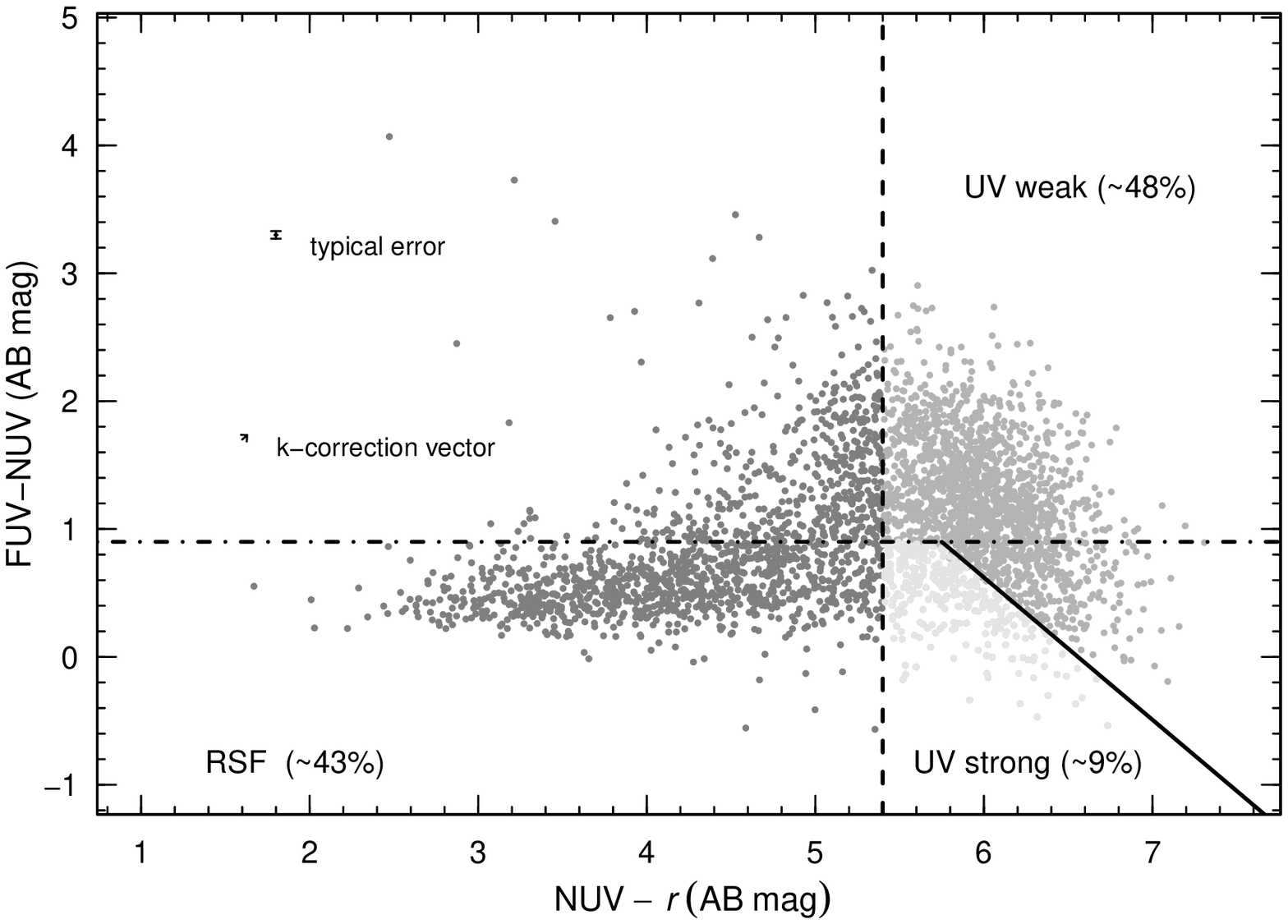}
\includegraphics[width=1.07\columnwidth,angle=0]{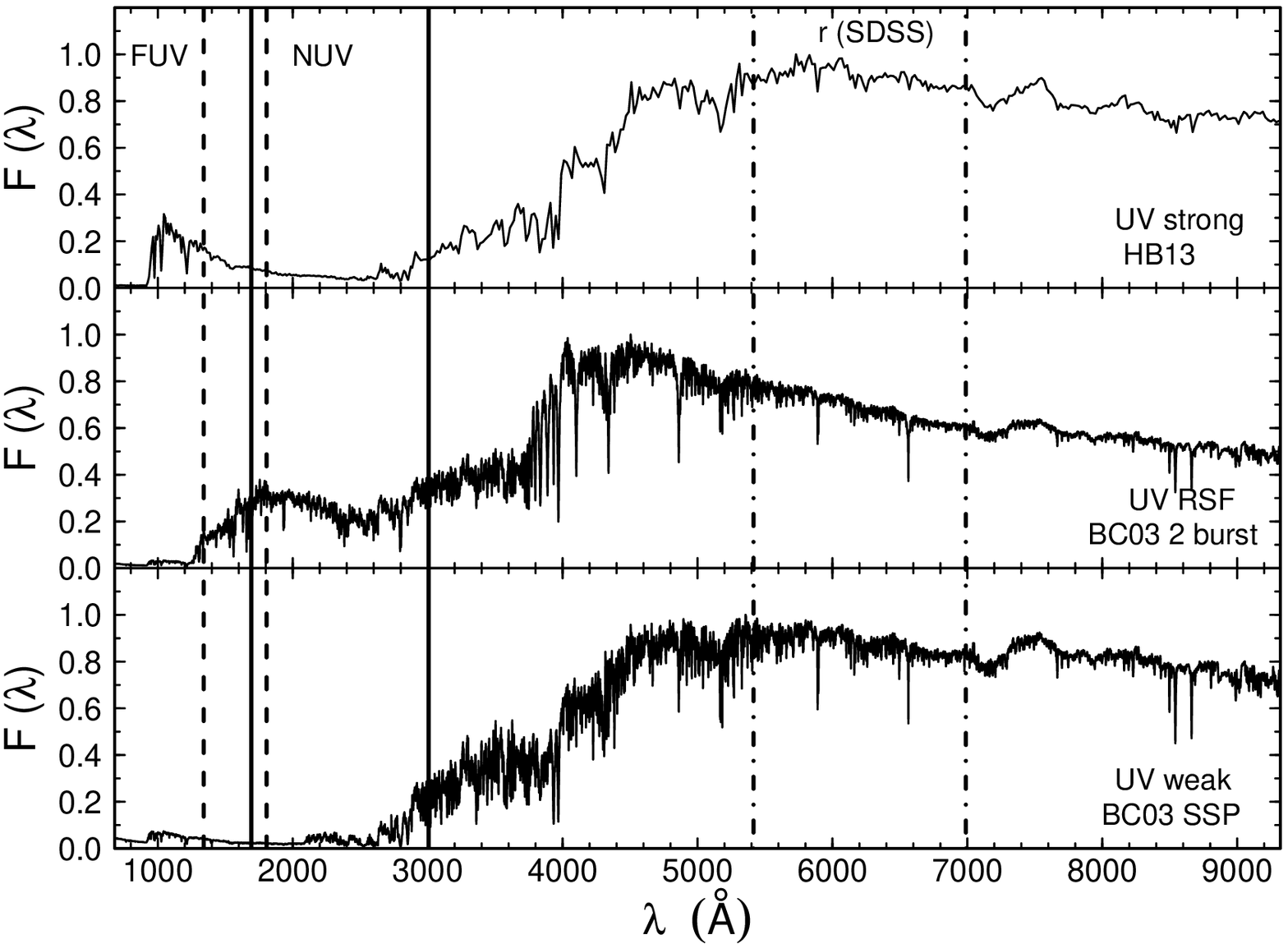}
\caption{
\textit{Top:} 
\uvr\ showing the 3417 galaxies classified as ETGs in \S~\ref{sec:sample}. 
Vertical dashed line: NUV-\textit{r} = 5.4, 
horizontal dot dashed line: FUV-NUV = 0.9, 
and inclined solid line: FUV-\textit{r} = 6.6
(see Table~\ref{tab:UV_esq}).
Typical error bars and \textit{k}-correction vector are plotted. The data has been corrected for Galactic extinction.
Red and black dots represent UV strong (UVX) and UV weak ETGs, respectively.
Blue dots correspond to ETGs with recent star formation. 
\textit{Bottom:} 
Typical spectral energy distribution of UV strong (UVX), RSF and UV weak ETGs.
The UVX spectrum was computed using the HB13 models. The BC03 models were used to compute the RSF and UV weak galaxy spectra.The wavelength range covered by the FUV, NUV, and $r$ filters are indicated by the vertical lines.}

\label{fig:dcc}
\end{figure}

The spectral energy distribution of many early type galaxies (ETGs) and spiral galaxy bulges increases to shorter wavelengths in the range from 1200 to 2000 {\AA}
\citep*{Code79}. 
This phenomenon is called UVX or UV upturn (cf. Figure~\ref{fig:dcc}).
Observational evidence shows that UVX increases with stellar metallicity.
\cite{Burstein88} found that ETGs with bluer 1550-V colour have larger stellar velocity dispersion $\sigma$ and higher absorption line index $\mathrm{Mg}_2$.
Giant elliptical galaxies with large $\sigma$ values have [Mg/Fe] index higher than most metal rich stars in the solar neighbourhood \citep*{WFG92}.
The work of \cite{Boselli05} and \cite{Donas07} reinforces the conclusion of \cite{Burstein88} that metallicity is a fundamental parameter, 
and that the stronger cases of UVX occur in massive metal rich ($Z > Z_{\odot}$) ETGs.

The launch of the Galaxy Evolution Explorer (\galex) satellite revolutionised the study of the UV emission of galaxies \citep{Martin05}.
Combined \galex\ and Sloan Digital Sky Survey (\sdss, \citealt{Stoughton02}) observations have revealed that a substantial fraction ($\sim$ 30\%) of ETGs
exhibit strong NUV and FUV emission, consistent with residual star formation \citep{Kaviraj07,Schawinski07}, and that only a small percentage ($\sim$ 5\%)
of ETGs shows the classical UVX \citep{Yi05,Yi11}.

Using GALEX colours combined with SAURON spectroscopy, and rejecting galaxies with recent star formation, \cite{Bureau11} found that galaxies
with bluer FUV-V colour have stronger Mg\textit{b} and lower $\mathrm{H}\beta$ absorption.
This result suggests that galaxies with bluer FUV-V are older or have higher metallicity, or both.
A similar conclusion was reached by \cite*{Smith12} studying a sample of 150 red and passive Coma cluster galaxies
with spectroscopic age and estimated element abundances determinations.
They also found a strong correlation of FUV-NUV colour with age and metallicity.

Nowadays, the favoured candidate sources for the UVX are extreme horizontal branch (EHB) stars and their descendants \citep*{GR90,Rosenfield12}. 
EHB stars are low-luminosity core He-burning stars with temperatures reaching up to 23000 K and lifetime $\sim 10^8$ yr.
Hence, they are difficult to observe, but they contribute considerably in the UV region of the spectrum, 
especially in galaxies dominated by old stellar populations like ETGs.
EHB stars are observed in galactic globular and open clusters (e.g. NGC2808, $\omega$ Cen, NGC 6791) and among field stars \citep{Catelan09}.
The dwarf elliptical galaxy M32 \citep{Brown00} is the only extragalactic stellar system with resolved EHBs.   
\cite{Rosenfield12} observed over 4000 old hot stars in the bulge of M31 and classified them into three classes: 
Post-AGB (P-AGB) stars, Post-Early AGB (PE-AGB) stars, and AGB-manqu\`e stars.
They conclude that the stars detected contribute only about 2\% of the total flux in the WFC3/UVIS F275W
filter, and suggest that the missing stars responsible for this emission are most likely faint EHB stars.
This result indicates that the stars responsible for the UVX have not yet been detected with certainty.
 
A number of competing mechanisms have been proposed to explain the origin of EHB stars.  
\cite{GR90} argue for a single star origin. The large amount of mass loss during the RGB phase of high metallicity stars
leads to a hot, low envelope mass EHB star. Once the star leaves the HB, depending on the envelope mass and the metallicity,
the EHB star may skip the AGB phase and evolve through a more luminous post-EHB phase: either P-AGB, or PE-AGB, or AGB-manqu\`e.
The presence of EHB stars and the strength of the UVX should then depend strongly on the age and metallicity of the stellar population.
On the other hand, \cite{Han02,Han03} propose that EHB stars result from the interaction of binary stars through one of three possible channels:
common-envelope (CE) ejection, stable Roche lobe overflow (RLOF), or two Helium white dwarf merger (2HeWD).
In this scenario, the strength of the UV upturn should depend weakly on the age and metallicity of the stellar population \citep*{Han07,Han10}.

One motivation of the work by \cite{Smith12} was to distinguish between the single or binary star origin of EHB stars using the UVX in passive ETGs.
They favour the metal rich single star origin of these stars.
The main argument against the binary star origin of EHB is the expectation that, at least to some level, the UVX should be a universal phenomenon
observable in all ETGs, uncorrelated with the age and metallicity of the stellar population.

In this paper we explore to what extent stellar population synthesis models that include the formation of EHB stars through binary star interactions
can explain the UVX in ETGs.
In \cite*{\hpb} (hereafter, \hpb) we presented such a model and showed that it can explain successfully the observed colour-magnitude diagram
of the metal-rich galactic open cluster NGC 6791. Confirmed EHB stars are present in this cluster.
If we consider the stellar population in this cluster archetypal of the stellar population in ETGs that show UVX, then our model should be 
adequate to study the UV upturn in ETGs.
In \S~\ref{sec:sample} we assemble a sample of ETGs with UV and optical detections, and
use the UV$r$ colour-colour plane to classify these galaxies as either star forming, UV weak, or UVX.
In \S~\ref{sec:model} we review the most important aspects and parameters of the \hpb\ models.
The predicted UV-optical colours from these models are compared with the data in \S~\ref{sec:ModelPredictions}.
The results are discussed in \S~\ref{sec:discussion} and the conclusions are presented in \S~\ref{sec:conclusion}. 

\section{The ETG sample}
\label{sec:sample}

To study the UVX phenomenon in ETGs using the \hpb\ models we must build a sample of galaxies observed both in the UV and optical ranges.
The \sdss\ and the \textit{GALEX Medium Imaging Survey} (\galexmis) overlap over 1000 square degrees in the sky and contain a considerable number of objects in common.
We use the following \textit{SDSS} pipeline parameters to select ETG candidates from the \sdss-DR8\footnote
{The DR8 (the eighth \sdss\ data release) is complete to \textit{r} = 22.2 mag.
$fracDev\_x$ is a structural parameter which indicates the fraction of the light of a galaxy that follows the \cite{Vaucouleurs} $\mathrm{r}^{1/4}$ 
law in the $x$ band. For the \textit{r} band, $\lambda_{eff}$ = 6165 \AA.}
\citep{DR8a,DR8b}, which will then be matched to their \galex\ counterparts.

\begin{enumerate}[(i)]
\item $fracDev\_g >$ 0.95: rejects spiral arms and blue disks,
\item $fracDev\_r >$ 0.95: selects objects that follow the $\mathrm{r}^{1/4}$ law in the $r$ band, and
\item  $fracDev\_i >$ 0.95: reinforces constrain (i)
\end{enumerate}  

These purely morphological criteria, with no assumption on colours or spectral energy distributions, were proposed by \cite{Kaviraj07} and \cite{Schawinski07}
in order to include more genuine ETGs than, e.g., \cite{Bernardi03}.
According to \cite{Kaviraj07} and \cite{Schawinski07} these criteria are fulfilled by ETGs brighter than $r = 16$.
Selecting galaxies with $r \le 16$ guarantees the completeness of the sample at the top of the red region (NUV-\textit{r} = 6.2) defined by \cite{Yi05}. 
Additionally, we keep only those galaxies with good spectral quality (S/N $>$ 10) and redshift \textit{z} $<$ 0.1.
This limits the volume of the sample to the local universe, so that no cosmological considerations enter this study. 
We end up with a sample of 64142 \sdss\ local ETGs brighter than $r = 16$.

Now we proceed to find the \sdss\ ETG candidates present in the \galex\ sample as well.
The \textit{GALEX-MIS} sixth data release (GR6) contains \mbox{1500 s} single orbit exposures in the NUV band (near-ultraviolet, centred at 2271 \AA) and the 
FUV band (far-ultraviolet, centred at 1528 \AA).
The AB-system limiting magnitude in these bands is 22.7 and 22.6, respectively \citep{Morrissey05}.
\cite{Budavari09} applied Bayesian statistics to achieve a cross-matched catalog using a large amount of data from the \sdss\ and the \textit{GALEX} surveys. 
They found that a search radius of 4 arc sec in the ($\alpha, \delta$)\footnote{Right Ascension and Declination}
source position is acceptable to match objects present in both surveys. We adopt their value in this study. 
By means of SQL queries to \textit{GALEX} CasJobs, we selected 5771 \sdss\ ETGS detected in both the FUV and the NUV bands
with S/N $\gtrsim$ 3 in the NUV.

Active Galactic Nuclei (AGN) are present in our sample.
AGN in the local universe host  preferentially massive elliptical galaxies \citep{Kauffmman03}.
Type I AGN (Quasars) are easily removed from the sample using the tag included in the \sdss\ pipeline.
To remove type II AGN (Seyfert and LINERs) and star bursts (SB) we use the line ratios \mbox{[OIII]/$\mathrm{H}_{\beta}$} and \mbox{[NII]/$\mathrm{H}_{\alpha}$},
as proposed by \cite{Kauffmman03}, in a modified version of the BPT diagram \citep*{BPT}:

\begin{enumerate}[(i)]
\item Seyfert: [NII]/$\mathrm{H}_{\alpha}>0.6$ and [OIII]/$\mathrm{H}_{\beta} > 3$,
\item LINER:  [NII]/$\mathrm{H}_{\alpha}>0.6$ and [OIII]/$\mathrm{H}_{\beta} < 3$, and
\item SB: $log([\mathrm{OIII}]/\mathrm{H}_{\beta})>0.61/\{log([\mathrm{NII}]/\mathrm{H}_{\alpha}-0.05\}+1.3$.
\end{enumerate}

\noindent
According to these rules, $\sim$ 26\% of the galaxies in our sample are AGN.
This is consistent with the $\sim$ 25\% of type II AGN found by \cite{Kaviraj07} in their ETG sample.

Even though the \cite{Kaviraj07} and \cite{Schawinski07} morphological classification excludes late-type galaxies, and the BPT method removes possible intruders,
an important fraction of ETGs may have undergone recent star formation.
Using the \cite*{bc03} stellar population synthesis models (BC03 hereafter) computed for the Stelib \citep{jfl03}, Miles \citep{ps06}, and IndoUS  \citep{fv04} stellar
libraries over the range of typical ETG solar and super-solar metallicity, we find that for a simple stellar population (SSP) in the age range 5-10 Gyr, the 
equivalent width (EW) of $\mathrm{H}_{\beta} \, \lesssim$ 2.1 \AA. A similar result was found by \cite{Bureau11}. 
Discarding galaxies with EW $\mathrm{H}_{\beta} > \, $ 2.1 \AA, we end up with a catalogue of 3417 ETGs with both UV \galex\ and optical \sdss\ 
photometry, and no signs of nuclear activity and/or important events of recent star formation. 
In \S~\ref{sec:ModelPredictions} we will use this sample to compare with the predictions of the \hpb\ models.

\begin{table}
\centering
  \caption{UV classification scheme \citep{Yi11}.} \label{tab:UV_esq}
  \begin{tabular}{cc}
  \hline
  Criterion                           &   Description  \\
 \hline
FUV-NUV $<$ 0.9          &  UV rising slope. \\ 
NUV-\textit{r}  $>$ 5.4    &  Devoid of young massive stars. \\
FUV-\textit{r} $<$ 6.6     &  Strong FUV flux (UVX).  \\ 
\hline
\end{tabular}
\end{table}

\subsection{UV-classification scheme}
\label{sec:UVscheme}

The position of the 3417 ETGs in the (FUV-NUV) vs. (NUV-$r$) two colour diagram (\uvr\ hereafter) is shown in Figure~\ref{fig:dcc}.
The colours of the galaxies shown in the Figure have been corrected for Galactic foreground extinction using the \cite*{Schlegel98} maps.
We assume the UV extinction law from \cite{Wyder05}, and the \cite{Fukugita04} formula for the \sdss\ \textit{r} band.
The galaxies in our sample are in the local universe (z $\lesssim$ 0.1), therefore, evolutionary corrections are irrelevant.
In Figure~\ref{fig:dcc} the arrow indicates the typical \textit{k}-correction vector determined by \cite{Yi11}.
The length of this arrow is similar to the size of the dots representing galaxies, hence the \textit{k}-correction is negligible.

The wide range of (FUV-NUV) and (NUV-$r$) colours covered by the ETGs in Figure~\ref{fig:dcc} indicates 
that the relative amount of UV flux in these galaxies is far from uniform.
This behaviour can be explained by the presence of either EHB stars, hot evolved stars, or young massive stars in different proportions in each galaxy.
The slope of the UV spectrum is related to the temperature of the stars emitting in the UV.
\cite{Yi11} established a colour criterion to classify ETGs according to their UV spectral morphology, illustrated in Figure~\ref{fig:dcc} and summarised
in Table~\ref{tab:UV_esq}. 
FUV-NUV = 0.9 indicates a flat UV spectrum, whereas a bluer FUV-NUV corresponds to a steeper slope.
Passively evolving ETGs have NUV-\textit{r} $\ge$ 5.4 and FUV-\textit{r} $>$ 6.6.
The NUV-\textit{r} = 5.4 line indicates the blue limit of the red sequence.
UV strong ETGs obey the three conditions in Table~\ref{tab:UV_esq}.

As indicated in Figure~\ref{fig:dcc}, according to the \cite{Yi11} UV classification criterion,
$\sim$ 48\% of the galaxies in our sample of ETGs are classified as UV weak galaxies,
$\sim$ 43\% as residual star forming (RSF) galaxies, and only
$\sim$   9\% as UV strong (or UVX) galaxies. 
These percentages are in good agreement with those found by \cite{Yi11} only if galaxies detected in both the 
FUV and the NUV are considered, as we do in our sample\footnote{Our sample of ETG's will be made available to the interested reader. See Table~2}.
 Our models do not intend nor pretend to explain the percentage of galaxies seen in each region of Figure~\ref{fig:dcc}.
UVX is present in about 20\% of non RSF ETG's. It is likely that UVX is also present in a similar fraction of RSF galaxies,
but is hidden by the UV emission of young massive main sequence stars present in these galaxies.
In the next section we discuss some scenarios that may explain why 80\% of the ETG's avoid the UVX phenomenon.

\begin{figure*}
\begin{center}
\includegraphics[width=2.08\columnwidth,angle=0]{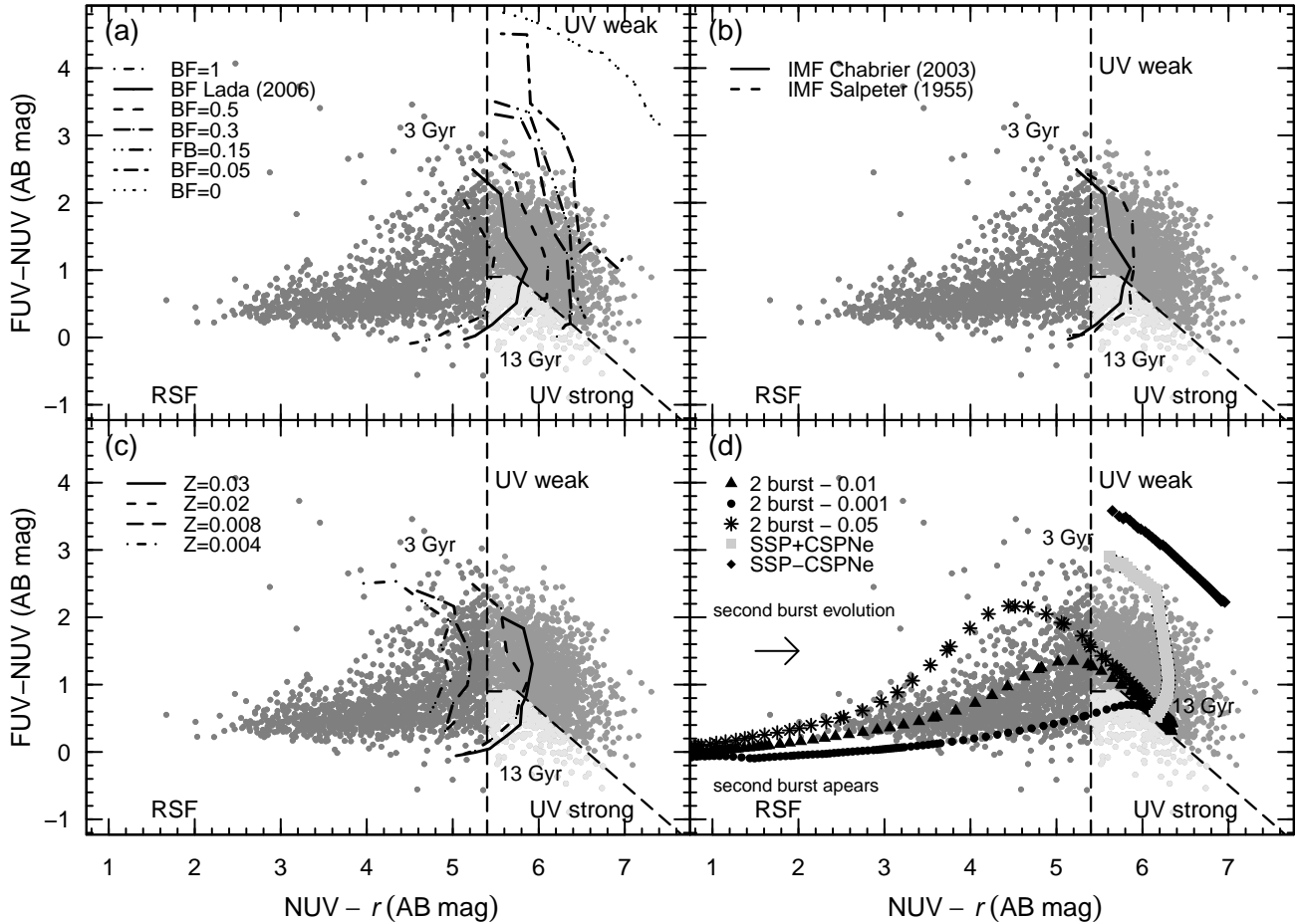}
\caption{
\textit{a,b,c:} 
The lines show the colour evolution of various \hpb\ models in the \uvr.
Time varies from 3 Gyr at the red FUV-NUV end of the line to 13 Gyr, where models reach the region of UV strong (UVX) ETGs.
$(a)$ Colour evolution for different values of the fraction of binary stars (BF) as indicated in the figure.
$(b)$ Colour evolution for the \citet{Chabrier03} and \citet{Salpeter55} IMF models.
$(c)$ Colour evolution for models of different stellar metallicity $Z$, as indicated in the figure.
$(d)$ Colour evolution of standard BC03 models in which a second burst of star formation occurs when the stars formed in the first burst are 12 Gyr old.
Different symbols represent bursts of different amplitude, indicated in the figure as the fraction of the total mass.
These bursts add 0.1, 1, and 5\% to the initial mass of the galaxy.
When the second burst occurs the NUV-\textit{r} colour changes drastically towards the blue, reaching the NUV-$\textit{r} < 3$ RSF region in the \uvr.
As the massive stars in the second burst fade away, NUV-\textit{r} becomes redder and the stellar population eventually recovers the colours it had before the second burst.
The higher the amplitude of the second burst, the redder the FUV-NUV colour reached by the stellar population in the post burst era.
For reference, a BC03 SSP model, including and excluding CSPNe, is also plotted.
}
\label{fig:mosaic}
\end{center}
\end{figure*}

\section{SPS including binary stars}\label{sec:model}

In \hpb\ we constructed a stellar population synthesis (SPS) model that includes interacting binary stars.
We use the \cite*{Hurley02} BSE public code to compute the evolutionary tracks of a population of binary pairs.
The BSE code follows the evolution of the physical and orbital parameters of both members of a pair
through all phases of stellar evolution from the ZAMS up to the remnant stage.
The distributions of period, mass ratio, and the fraction of binary stars are carefully modelled to reproduce the observational trends as much as possible.
The most relevant binary star interactions, including RLOF, CE, mass transfer, mass accretion, supernova kicks, and angular momentum loss
are followed in the BSE code.
The occurrence of one or more of these processes depends on the orbital and initial physical parameters of the pair.
In the standard version of the BSE code, when the 2HeWD in a binary pair collide, the temperature of the merger product becomes
too hot and the object is destroyed. In our version of the BSE code, we use the \cite{Han02} conditions for He to ignite after the 2HeWD merger occurs.
In this case, we assign physical parameters to the resulting EHB star using the BaSTI data set, based on evolutionary tracks computed with realistic stellar physics by \cite{Adriano04,Adriano06}. Our code then includes EHB stars formed through one of these channels: RLOF, or CE, or 2HeWD merger.

In \hpb\ we used our model to explain successfully the observed colour-magnitude diagram of the metal-rich ($Z = 0.03$) galactic open cluster NGC 6791,
which contains a significant number of confirmed EHB stars, resulting in a synthetic spectrum with a strong UVX.
If the stellar population in this cluster is archetypal of the stellar population present in UVX galaxies, then our model should be 
adequate to study the UV upturn in ETGs. See \hpb\ for details.

\section{Modelling the UVX}\label{sec:ModelPredictions}

\subsection{Model parameters}\label{sec:modparam}

Observational evidence shows that ETGs exhibit different degrees of UV emission (cf. Figure~\ref{fig:dcc}).
In this section we examine the range of values covered by the UV  colours and UVX as we vary the \hpb\ model input parameters.
In these models the rate of EHB star formation is controlled by the following parameters: 
(\textit{a}) Fraction of binary stars (BF),
(\textit{b}) Initial mass function (IMF), and
(\textit{c}) Stellar metallicity ($Z$).
Unless stated otherwise, models discussed below represent SSPs computed for the \cite{Chabrier03} IMF,
the \cite{Lada06} spectral type dependent BF, and the solar metallicity ($Z = 0.02$) \cite{Hurley02} binary star evolutionary tracks.
The IMF extends from 0.1 to 100 $\textrm{M}_{\odot}$.

Additionally, a binary pair is characterised by its orbital period ($P$), orbit eccentricity ($e$), and mass ratio ($q = M_2 / M_1$).
As in \hpb\ we assume that $e$ and $q$ are distributed uniformly, following \cite{Zhang04,Zhang05}, and \cite{Milone12}, respectively,
and that $P$ follows the Gaussian distribution (in \textit{log\,P}) found by \cite{DM91},
with $\overline{log\, P}$ = 4.4, and \mbox{$\sigma_{log\, P}$ = 2.3}, where \textit{P} is measured in days.
In the tests below the distributions of $P$, $e$, and $q$ are kept fixed.

\subsubsection{Binary Fraction} \label{sec:fb}

Since in the \hpb\ models all EHB stars are formed via binary star interactions,
the resulting number of EHB stars depends on the value of BF characterising the population.
To study the influence of BF on the UV colours of the model population, we computed \hpb\ models for the following values of BF:
(\textit{a}) BF=0, all stars are single; 
(\textit{b}) BF=0.05, 5\% of the stars are in binary systems;
(\textit{b}) BF=0.15, 15\% of the stars are in binary systems;
(\textit{b}) BF=0.3, 30\% of the stars are in binary systems; 
(\textit{c}) BF=0.5, 50\% of the stars are in binary systems;
(\textit{d}) BF=Lada, BF depends on spectral type following \cite{Lada06}; and
(\textit{e}) BF=1, all stars are in binary systems.
The predicted colour evolution is shown by the different lines in Figure~\ref{fig:mosaic}a.
Models with BF $\ge 0.30$ reproduce the observed colours of ETGs in both the UV-weak and UV-strong regions of the \uvr.
It is significant that a minimum BF is required to reproduce these colours.
The colours of ETGs catalogued as RSF cannot be reproduced, not even for BF = 1.

\cite{DOR95} developed a spectral synthesis model that includes all phases of single star evolution, from the ZAMS to the white dwarf phase. 
They find that for the bulk of ETGs $\sim$ 5\% of the stars on the HB are EHB stars, 
whereas for UVX galaxies this fraction reaches $\sim$ 20\%, which is not far from our results. 
\cite{Kalirai07} determined that at least 40\% of the evolved stars in NGC 6791 have lost enough mass on the RGB phase to avoid the helium flash, and
\cite{Bedin08} established that in NGC 6791 BF $\sim$ 25-35\%. 
These results suggest that BF is not the same in all stellar systems.

\subsubsection{Initial Mass Function} \label{sec:imf}

The rate of production of atypical stars by interacting binary pairs, like EHB stars or blue stragglers, depends on the frequency 
of progenitors in the appropriate mass range, which is determined by the IMF. 
The probability to form a binary system that will evolve into two HeWD will be higher for those IMFs richer in their progenitors.
Thus, despite the narrow mass range (0.4-0.65 $\textrm{M}_{\odot}$) of the progenitors of EHB stars formed through the 2HeWD merger channel,
we expect a slight dependence of the number of these stars on the IMF.

Figure~\ref{fig:mosaic}b shows the colour evolution of models computed with the \cite{Salpeter55} and the \cite{Chabrier03} IMFs.
For both IMF's the models reproduce the range of colours of passively evolving ETGs in the UV weak and UV strong regions of the \uvr. 
The Chabrier IMF model colours are bluer in NUV-\textit{r}. 
The frequency of EHB progenitors is higher for the Chabrier than for the Salpeter IMF.
The Salpeter IMF yields a higher number of stars with mass below 1 $\textrm{M}_{\odot}$ than the Chabrier IMF. 

\subsubsection{Metallicity} \label{sec:z}

The number of EHB formed by binary interactions should not depend on stellar metallicity, but only
on the distribution of the orbital parameters of the pairs \citep{Han07}.
\cite{Smith12} argue that the influence of metallicity on these models is neither completely addressed nor fully evaluated,
because even if the formation of the binary systems does not depend on metallicity, the subsequent evolution of the EHB star does.
The physical parameters of EHB stars in our models are derived from the BaSTI data set, which does include the metallicity dependence.
Figure~\ref{fig:mosaic}c shows the colour evolution of models computed for $Z = 0.004, 0.008, 0.02$, and 0.03.
As expected, high $Z$ models ($Z = 0.03$ and 0.02) reproduce well the colours of UV weak and UV strong ETGs.
The  $Z =0.03$ model is slightly bluer than the $Z = 0.02$ model.
Thus, the \hpb\ models predict a correlation between the strength of UVX and metallicity.
The \cite{Hurley02} code does not allow to compute tracks for $\bf{Z > 0.03}$.
Subsolar models ($Z = 0.008$ and 0.004) are bluer in (NUV-\textit{r}) because of the presence of blue HB stars resulting from the evolution of single stars. 

\subsection{Two bursts of star formation} 
\label{sec:2burst} 

Although most of the ETGs in our sample are located in the UV weak and UV strong regions of the \uvr, approximately 43\% of the galaxies are located in the RSF region.
\cite{Kaviraj07}, \cite{Schawinski07}, \cite{Kaviraj08}, and \cite{Kaviraj10} report observational evidence that the blue NUV-\textit{r} colour of these galaxies
is due to residual star formation within the last 1 Gyr.
In particular, \cite{Kaviraj10} concludes that ETGs with blue UV colours are morphologically perturbed and that these colours are driven by merger-induced
star formation within the last 3 Gyr.

Galaxies with NUV-$\textit{r} < 5.4$ can not be explained in terms of the \hpb\ SSP models discussed in the previous subsections 
and shown in Figures~\ref{fig:mosaic}a,b,c. 
\hpb\ models for arbitrary star formation histories are not yet available.
To understand the colours of the bluest galaxies in Figure~\ref{fig:mosaic}, we use the standard BC03 models.
The BC03 models do not include binary stars, and lack the EHB star contribution present in the \hpb\ models.
However, the massive main sequence O-B stars formed in a recent burst of star formation will outshine in the UV the contribution of the EHB formed in the initial burst.
In Figure~\ref{fig:mosaic}d we show the path followed in the \uvr\ by models in which a second burst of star formation occurs when the stars formed in the first burst are 12 Gyr old. Different symbols represent bursts of different amplitude, as indicated in the figure.

The standard BC03 SSP model is also shown in Figure~\ref{fig:mosaic}d. The central stars of planetary nebulae (CSPNe) included in this model are responsible of the UV flux seen at late ages. For comparison we include an SSP models in which CSPNe have been suppressed. This latter model
resembles the HB13 model for BF = 0.

The lowest amplitude burst (0.1\% ) traverses the UV strong region in Figure~\ref{fig:mosaic}d on its way to bluer colours.
However, the time spent by this model in the UV strong region is quite short, less than 0.15 Gyr. 
The effects of this burst on the integrated spectrum last for no longer than 0.7 Gyr. 
Therefore, the probability that all ETGs catalogued as UV strong have suffered a recent (in the last $\sim$ 0.55-0.7 Gyr) and low amplitude ($\sim$ 0.1\% of the total mass) burst is small.  Additionally, studies of resolved stars in the bulge of M31 \citep{Rosenfield12} and in M32 \citep{Brown00}, do not show the presence of MS stars.

\section{Discussion}\label{sec:discussion}

The principal aim of this study is to fill the gap between the classical models (single star scenario) to explain the UVX phenomenon \citep*[e.g.][]{Yi97} and models that include interacting binaries \citep[e.g.][]{Han10}.
In both type of models the UVX is produced by EHB stars (He burning stars with $T_{eff} >$ 23000 K).
EHB stars are considered the most likely source of the UVX despite the fact that they are not resolved in extragalactic objects \citep{Brown97,Brown08,Rosenfield12}.
In the classical model, EHBs are the product of the evolution of high $Z$ or He enhanced low mass single stars.
In the interacting binary models EHB stars are formed via binary interactions through mass transfer, collisions, or mergers. 
Therefore, the two models make different predictions for the evolution of the UVX in ETGs.
In the classical model, the intensity of the UV flux increases with the age of the stellar population because low mass stars have less gravitationally bound
lower mass envelopes, which are easier to lose through mass loss during the RGB evolution, producing blue HB stars.
Since the mass loss rate increases with $Z$, in this model the UVX is expected to increase with the metallicity of the stellar population.
\cite{Carter11} and \cite{Smith12} find such a correlation of UVX with age and metallicity, and conclude that the most probable source of the UVX is evolved
low mass stars. 

On the other hand, from interacting binary models we do not expect a relation between the UVX and age or $Z$ \citep{Han02}.
\cite{Han07} show that the FUV-\textit{r} colour is almost constant for age between 3 and 12 Gyr.
In principle, in these models the UVX is not related to $Z$ because the formation channel of EHBs depends only on the orbital parameters of the binary pairs.
But, as indicated in \S~\ref{sec:intro}, once the EHB star is formed, its position and subsequent evolution in the HR diagram \textit{does} depend on $Z$.
The \hpb\ model predicts that the ETGs inside the UV strong region are the oldest (Figure~\ref{fig:mosaic}). 
This result is related to the use of the \cite{Adriano04,Adriano06} evolutionary tracks to assign the physical properties to EHBs formed as a result of a 2HeWD merger.
These tracks allow for a narrow range of mass for the EHB stars, form 0.49 to 0.53 $\textrm{M}_{\odot}$. 
Since the mass of the EHB star is the sum of the mass of the two merging WDs, the probability of producing an EHB star in the indicated mass range
via the 2HeWD merger channel increases with time.
As shown in Figures~\ref{fig:mosaic}a,b,c, in the \hpb\ models the UVX phenomenon appears in old stellar populations  ($\gtrsim$ 7 Gyr).
Figure~\ref{fig:mosaic}c shows a weak correlation between $Z$ and the UV colours inside the UV strong region of the \uvr.
The $Z = 0.03$ model is slightly bluer that the $Z = 0.02$ model in the UV strong region.
A more detailed treatment of the evolution in the post-EHB phases would help establishing the reality of these two correlations.

It is important to remark that the \hpb\ models reproduce all ETGs in the sample catalogued as UV weak or UV strong for some choice of BF, $Z$, or the IMF.
The distribution of these parameters, and others like orbital period and mass ratio, and its temporal evolution and dependence with environment 
remain an unresolved puzzle. 

\section{Summary and Conclusions}
\label{sec:conclusion}

We use the \hpb\ SPS models, which include the evolution of interacting binary stars, as a tool to study the UVX phenomenon observed in ETGs.
First we build a sample of ETGs with observed optical (\textit{SDSS}-DR8) and UV (\textit{GALEX}-GR6) photometry.
Following \cite{Yi11}, we classify the ETGs in our sample according to their position in the \uvr\ as UV weak (galaxies in the red sequence), 
UV strong (galaxies with UVX), and RSF (galaxies that have experienced recent episodes of star formation).
Then, we compare the UV colours of the sample galaxies with the colour evolution lines in this plane predicted by the \hpb\ models, varying,
one at a time, the fraction BF of binary stars, the IMF, and the stellar metallicity $Z$.

In view of the results discussed in \S~\ref{sec:ModelPredictions}, we conclude that:

\begin{itemize}

\item Of all parameters tested, the UVX is most sensitive to BF. 
Varying BF it is possible to reproduce the UV colours of almost all the ETGs in the UV weak and UV strong regions of the \uvr,
as long as BF is above a threshold value of $\sim$ 0.15.

\item  Although the \cite{Chabrier03} IMF model is slightly bluer in NUV-\textit{r} than the \cite{Salpeter55} IMF model, both models are in good agreement
with the observed colours of UV weak and UV strong ETGs. 

\item Higher metallicity models ($Z = 0.02$ and $Z = 0.03$) reproduce the colours of UV weak and UV strong galaxies better than low $Z$ models.
The $Z = 0.03$ model is slightly bluer than the $Z = 0.02$ model in the UV strong region.
This behaviour seems indicative of a weak relationship between UVX and $Z$.
The HB stars present in the low $Z$ models move the integrated NUV-\textit{r} colour into the RSF region.

\item The colours of the ETGs in the RSF region of the \uvr\ cannot be understood in terms of the \hpb\ SSP models.
These colours are well reproduced by standard BC03 models, which lack binary stars, when a second burst of star formation is added to the old stellar population.
A burst strength of $\sim$ 5\% of the total mass of the stellar system suffices to explain the colours of the RSF galaxies.

\item The largest values of UVX occur at the oldest age in the model population (13 Gyr).

\end{itemize}

We would like to stress our last conclusion because the lack of a relation between UVX and age and $Z$ in models that 
assume that EHB stars result from binary interactions has been taken as a weakness of these models. 
Despite \cite{Smith12} remark that these models cannot account naturally for the observed trends of UVX with age and metallicity,
we have shown that the \hpb\ models predict a clear dependence of UVX with age, and to a lesser extent with $Z$.

When the tools to compute the evolution of binary pairs are improved, especially a better coverage of the tracks in the $(Y,Z)$ plane,
and an adequate treatment of the post-EHB phases of stellar evolution, it will be possible to explore these trends in depth.
In real galaxies EHB stars may form both from single stars and from binary pairs.

\section*{Acknowledgments}

We are very grateful to the referee, Phil Rosenfield, who provided useful comments and suggestions that helped to improve this paper.
FHP acknowledges the hospitality of the UNAM Centro de Radioastronom\'ia y Astrof\'isica during the last stages of this investigation.
FHP acknowledges support from CIDA during her PhD thesis work partially reported in this paper.
GB acknowledges support for this work from the National Autonomous University of M\'exico, through grants IA102311 and IB102212-RR182212.
This work has made use of the BaSTI web tools.

\bsp

\label{lastpage}


\clearpage

 \begin{figure*}
 \includegraphics{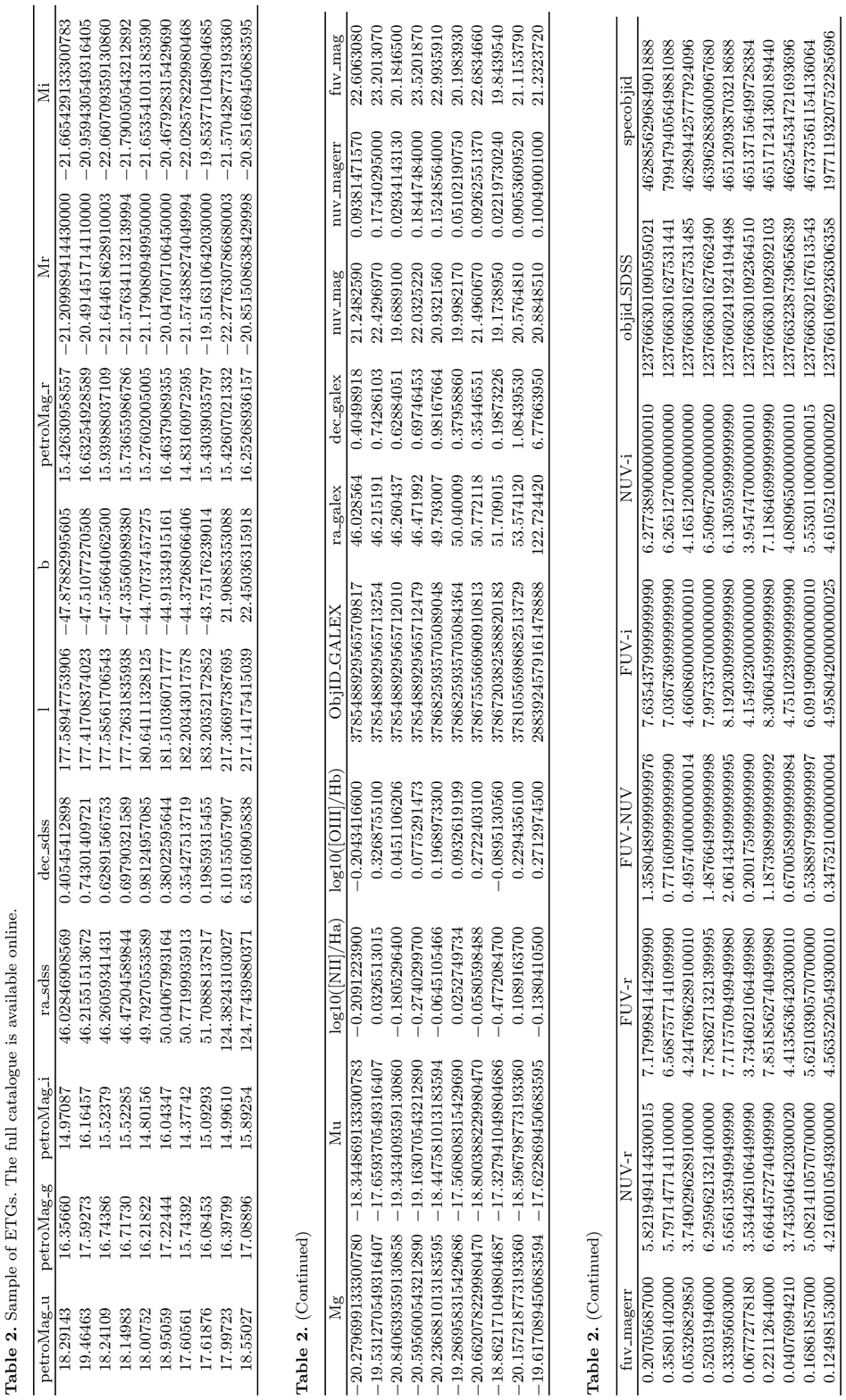}
 \label{}
\end{figure*}

\clearpage

 \begin{figure*}
 \hspace{-5cm}
 \includegraphics{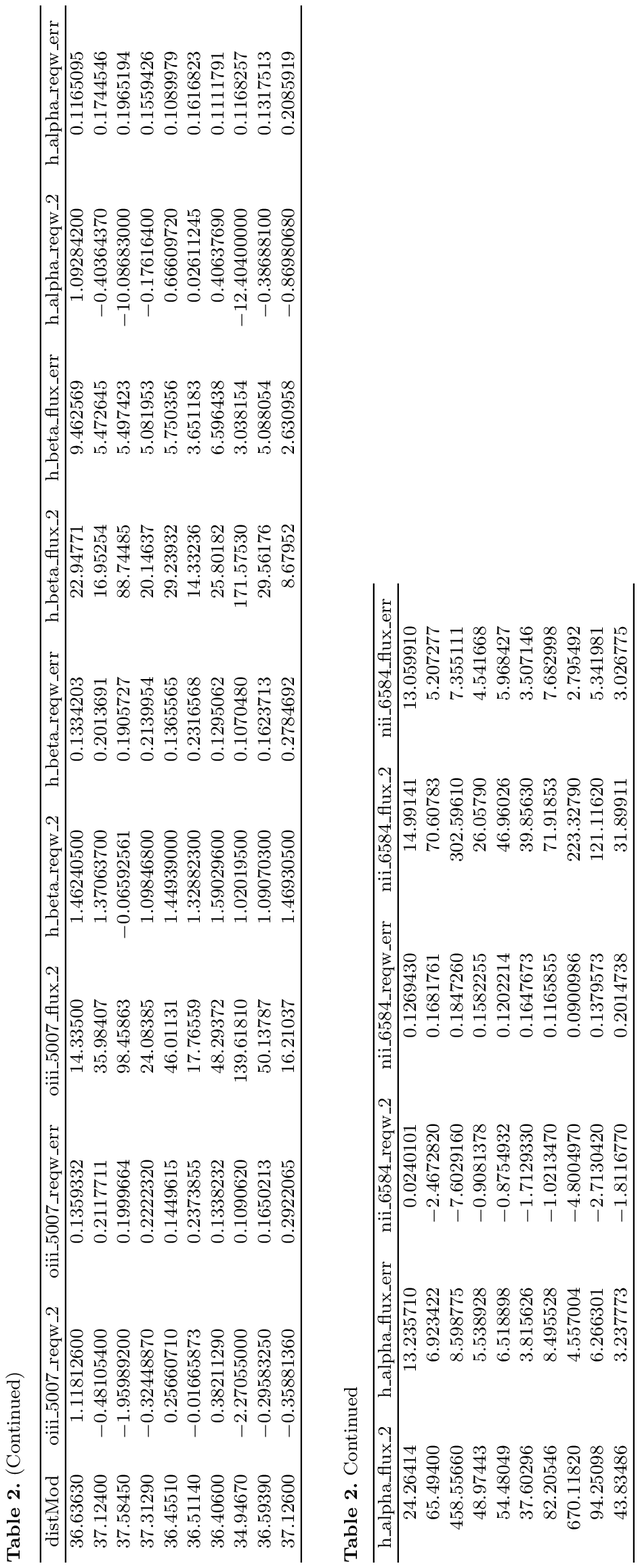}
 \label{}
\end{figure*}


\end{document}